\documentclass[11pt,a4paper,epsf]{article}

\newcommand{\ct}{\cite}
\newcommand{\lb}{\label}

\newcommand{\bc}{\begin{center}}
\newcommand{\ec}{\end{center}}
\newcommand{\bd}{\begin{displaymath}}
\newcommand{\ed}{\end{displaymath}}
\newcommand{\be}{\begin{equation}}
\newcommand{\ee}{\end{equation}}
\newcommand{\ba}{\begin{array}}
\newcommand{\ea}{\end{array}}
\newcommand{\bea}{\begin{eqnarray}}
\newcommand{\eea}{\end{eqnarray}}
\newcommand{\bt}{\begin{tabular}}
\newcommand{\et}{\end{tabular}}

\newcommand{\bp}{\begin{picture}}
\newcommand{\ep}{\end{picture}}
\newcommand{\bfi}{\begin{figure}}
\newcommand{\efi}{\end{figure}}

\newcommand{\eps}{\varepsilon}

\begin{document}
\vspace{0.5cm}
\bc
{\LARGE\bf The Effective QCD Lagrangian and Renormalization Group Approach.}\\
\vspace{0.4cm}
{\large\bf P.A.Kovalenko\footnote[1]{E-mail: pasha@heron.itep.ru},
L.V.Laperashvili\footnote[2]{E-mail: larisa@vitep5.itep.ru}}\\[0.4cm]
{\bf Institute of Theoretical and Experimental Physics,\\
B.Cheremushkinskaya 25, Moscow 117259, Russia}
\ec
\vspace{0cm}

\begin{abstract}
The perturbative part of the effective QCD Lagrangian was constructed
by the renormalization group approach in the 4-loop approximation.
It was extrapolated in the non-perturbative region under
the assumption of the existence of
gluon condensate at $(F_{\mu\nu}^j)^2=F_0^2$. The best result is
$F_0^2\approx 0.4$ GeV$^4$.
\end{abstract}

The present work is devoted to the construction of the
effective QCD Lagrangian which is controlled not only by perturbative (asymptotically free)
theory but also by non-perturbative strong coupling regime. The perturbative part of the QCD
Lagrangian is described here by the renormalization group (RG) approach \ct{1},\ct{2}.

Starting from the partition function for non-abelian SU(3)-gauge theory with quark fields
$\Psi$,$\bar\Psi$ and gluon fields $A_{\mu}^j$ (j=1,...,8 is the color index) we have the
following Euclidean functional integral:
\be
Z=\frac1N\int[{\cal D}A][{\cal D}\bar\Psi][{\cal D}\Psi]e^{-S(A,\bar\Psi,\Psi)}. \lb{1}
\ee

Assuming the appearance of the effective QCD action $S_{eff}$ as a result of the integration
over quark fields:
\be
Z=\frac1N\int[{\cal D}A]e^{-S_{eff}(A)}, \lb{2}
\ee
we have the following effective Lagrangian in Minkowski space:
\be
S_{eff}=\int d^4x {\cal L}_{eff}, \quad \quad
{\cal L}_{eff}=-\frac{\alpha^{-1}_{eff}(F^2)}{16\pi}F^2, \lb{4}
\ee
where $F^2\equiv(F_{\mu\nu}^j)^2$,\quad $ F_{\mu\nu}^j=\partial_{\mu}A_{\nu}^j -\partial_{\nu}A_{\mu}^j
+f^{jkl}A_{\mu}^kA_{\nu}^l$,\quad $\alpha_{eff}(F^2)=g_{eff}^2/{4\pi}$ and $g_{eff}$ is the effective
QCD coupling constant.

Below we consider the quantum corrections to the classical QCD action,
stipulated by the vacuum polarization induced not only by
quark loops, but also by gluon loops including ghosts. These contributions
don't change the form of the effective Lagrangian given by Eq. (\ref{4}).

The usual procedure used to estimate the generating functional:
\be
Z[J]=\frac1N\int[{\cal D}A]e^{-S_{eff}(A)+\int J_{\mu}^jA_{\mu}^jd^4x} \lb{5}
\ee
is to assume that the integral over all paths is dominated by minima of the
action $S_{eff}(A).$
For asymptotically free theories (including QCD) the minimum of $S_{eff}(A)$ is not the classical
vacuum $F_{\mu\nu}^j=0$. The quantum fluctuations in the strong coupling regime give rise to
spontaneous magnetization of the vacuum and the non-trivial vacuum field - the gluon condensate -
comes into existence: $F^2=F^2_0\ne0$. Using QCD sum rules the authors of Ref.\ct{3} have obtained
the following result:
\be
F^2_0\approx0.5 \mbox{ GeV}^4. \lb{6}
\ee
The effective action $S_{eff}$ gives the correct trace anomaly for the energy-momentum tensor
$\theta_{\mu\nu}$:
\be
\theta_{\mu}^{\mu}=\frac{\beta(g)}{2g^3}F^2, \lb{7}
\ee
where the effective coupling constant $g_{eff}\equiv g$ is a function of the variable:
\be
t=\frac12\ln\frac{F^2}{\mu^4_R}=\ln\frac{\mu^2}{\mu_R^2}, \lb{8}
\ee
($\mu$ is the energy scale and $\mu_R$ is the renormalization mass) and $\beta(g)$
is the well-known Callan-Symanzik $\beta$-function.

The effective action is RG invariant:
\be
\left[\mu^2\frac{\partial}{\partial\mu^2}+\beta(g)\frac{\partial}{\partial g}\right]S_{eff}=0,
\lb{9}
\ee
and $\alpha_s=g^2(t)/4\pi$ obeys the RG equation:
\be
\frac{d\ln\alpha_s(t)}{dt}=\beta(\alpha_s). \lb{10}
\ee

Considering the 4-loop approximation we have:
\be
\beta(\alpha_s)\approx-(\beta_0\tilde\alpha+\beta_1\tilde\alpha^2+\beta_2\tilde\alpha^3+
\beta_3\tilde\alpha^4) \lb{11}
\ee
where
$
\tilde\alpha\equiv\alpha_s/4\pi,\quad \beta_0=11-\frac23N_f, \quad \beta_1=102-\frac{38}{3}N_f,
$
$$
\beta_2=\frac{2857}{2}-\frac{5033}{18}N_f +\frac{325}{54}N_f^2,
\quad\quad \ct{4}
$$
$$
\ba{rcl}
\beta_3&=&\left(\frac{149753}{6}+3564\zeta_3\right)-
\left(\frac{1078361}{162}+\frac{6508}{27}\zeta_3 \right)N_f +\\
&&
\left(\frac{50065}{162}+\frac{6472}{81}\zeta_3 \right)N_f^2 +
\frac{1093}{729}N_f^3,\qquad \ct{5}
\ea
$$
Here $\zeta_s$ is the Riemann $\zeta-$function: $\zeta_3\approx 1.202056903...$ and $N_f$
is the number of quarks having masses $m_q<\mu$.

The solution of Eq.(\ref{10}) in the case of $3$-loop approximation gives
the following result:
\be
\ba{rcl}
\alpha^{-1}(t)&\approx&
\alpha_R^{-1}+\frac{\beta_0}{4\pi}t+
\frac{\lambda_1}{2}
\ln\frac
{\alpha_t^{-2}+\lambda_1\alpha_t^{-1}+\lambda_2^2}
{\alpha_R^{-2}+\lambda_1\alpha_R^{-1}+\lambda_2^2}+\\
&&\frac{\lambda_1^2-2\lambda_2^2}{\sqrt{4\lambda_2^2-\lambda_1}}
\arctan\frac
{\sqrt{4\lambda_2^2-\lambda_1^2}(1-\alpha_R\alpha_t^{-1})}
{(2+\lambda_1\alpha_R)\alpha_t^{-1}+\lambda_1+2\lambda_2^2\alpha_R},\\
\ea
\lb{12}
\ee
where we have used the following designations: $\alpha_t\equiv\alpha_s(t)$,
\quad $\alpha_R\equiv\alpha_s(\mu_R^2)$, \quad $\lambda_1=\frac{1}{4\pi}\frac{\beta_1}{\beta_0}$,
\quad $\lambda_2^2=\frac{1}{(4\pi)^2}\frac{\beta_2}{\beta_0}$.

We have also performed the computer solution of RG equation (\ref{10})
in the case of $4$-loop approximation. Starting from the value
\be
\alpha_s(M_Z)=0.118\pm0.003, \qquad \ct{6} \lb{13}
\ee
using the following (experimentally given) masses of $c$- and $b$-quarks \ct{6}:
\be
m_c=1.3\pm0.3 \mbox{\quad GeV}, \quad m_b=4.3\pm0.2 \mbox{\quad GeV}  \lb{14}
\ee
and varying the number of quarks $N_f$ from $N_f=3$ to $N_f=5$ we have reproduced the
function $\alpha_s^{-1}(t)$ in the perturbative region of energy scale $\mu\ge 1$ GeV
(see Fig.1 where the curves 1,..4 correspond to the n-loop approximations
with n = 1,..4). It was not difficult to construct the effective QCD
Lagrangian with the help of this behavior of $\alpha_s^{-1}(t)$.

The function $U=-{\cal L}_{eff}$ where ${\cal L}_{eff}$ is given by
Eq.(\ref{4}) coincides
with  the effective potential $U_{eff}$ only for constant $A_{\mu}^j(x)$
when $F_{\mu\nu}^j=f^{jkl}A_{\mu}^kA_{\nu}^l$. The perturbative function $U$ is presented
in Fig.2.

The general idea is to construct the curve $U(F^2)$ which, according to the confinement
scenario of Ref.\ct{7}, obeys the following requirements:
\begin{itemize}
\item $U(F^2)=0;$
\item $U(F^2)$ has the minimum at $F^2=F_0^2$;
\item the function $U(F^2)$ may be approximated by the Taylor expansion in
the vicinity of the minimum:
\end{itemize}
\be
U(F^2)\approx\sum_{n=0}^{3} a_n(F^2-F_0^2)^n. \lb{15}
\ee

The last approximation for $U(F^2)$ may be smootly sewn together with $U=-{\cal L}_{eff}$
derived from RG equation in the region $\alpha_s\sim 0.35$.

The result of such a procedure is given in Fig.3 and Table I.
\vspace{0cm}
\bc
$$
\ba{|c|ccc|cc|cc|c|}
\multicolumn{9}{c}{\mbox{\large\bf Table I.}}\\
\multicolumn{9}{c}{~}\\
\hline
\alpha_s&
\multicolumn{3}{|c|}{\mbox{condensate}}&
\multicolumn{5}{|c|}{\frac{i-\mbox{th loop}}{j-\mbox{th loop}},\% }\\
\cline{5-9}
\mbox{at sewing}&\multicolumn{3}{|c|}{ F_0^2, \mbox{GeV}^4}& 4/3 &&3/2&&2/1\\
\mbox{point}&+\sigma&\mbox{central}&-\sigma&&4/1&&3/1&\\
\hline
0.33 & 0.39 & 0.21 & 0.11 & 52\% &       & 21\% &       & 16\% \\
     &      &      &      &      & 1.7\% &      & 3.4\% &     \\
\hline
0.35 & 0.28 & 0.15 & 0.08 & 55\% &       & 22\% &       & 17.2\% \\
     &      &      &      &      & 2\%   &      & 3.8\% &        \\
\hline
0.4  & 0.16 & 0.08 & 0.04 & 60\% &       & 32\% &       & 23\%   \\
     &      &      &      &      & 4.3\% &      & 7.2\% &        \\
\hline
\ea
$$
\ec
The designation "central $\pm\sigma$" means:
\be
\ba{ccccc}
\left(
\ba{c}
\alpha_s\\
m_c\\
m_b
\ea
\right)
&=&
\left(
\ba{c}
0.118\\
1.3\\
4.3
\ea
\right)
&\pm&
\left(
\ba{c}
0.003\\
0.3\\
0.2
\ea
\right)
\ea
\lb{16}
\ee

The case "central $+\sigma$" gives the gluon condensate value close to the
result of Ref.\ct{3} : $F_0^2\approx 0.39$ GeV$^4$ (compare with Eq.(\ref{6})).
The corresponding curves for $U(F^2)$ and $\alpha_s^{-1}(t)$ are presented
in Fig.3 and Fig.4.
The function plotted in Fig.4 describes the behaviour of
$\alpha_s^{-1}(t)$
not only in the perturbative, but also in the non-perturbative regions showing
$\alpha_s(F_0^2\approx 0.39\mbox{ GeV}^4
\Leftrightarrow \mu\approx 0.79\mbox{ GeV})\approx -0.36$.

The equations of motion following from the effective Lagrangian (\ref{4}):
\be
{\cal D}_{\lambda}(\eps F_{\lambda\mu}^i)=j_{\mu}^i \qquad (i=1,...8)
\lb{17}
\ee
(${\cal D}_{\lambda}$ is a covariant derivative and $j_\mu$ is a current)
contain the chromodielectric function $\eps(t)$:
\be
\eps(t)=\frac{1}{4\pi}\left[\frac{d\alpha^{-1}}{dt}+2\alpha^{-1}(t) \right].
\lb{18}
\ee

It is not difficult to construct the function $\eps (t)$ with the help of the
function $\alpha_s(t)$ plotted in Fig.4. The preliminary
investigation of solutions of the equations of motion (\ref{17})
begins to show the existence of QCD strings.

We thank V.V.Vladimirsky and R.B.Nevzorov for useful discussions. The authors
also aknowledge the financial support of INTAS grants:
INTAS-93-3316-ext and INTAS-RFBR-95-0567.

\begin{figure}[t]
\bc
\bp(135,90)
%\put(0,-20){\special{psfile=gr8proc.ps hoffset=20
%\put(0,-20){\special{psfile=test1.ps hoffset=20
\put(0,-20){\includegraphics{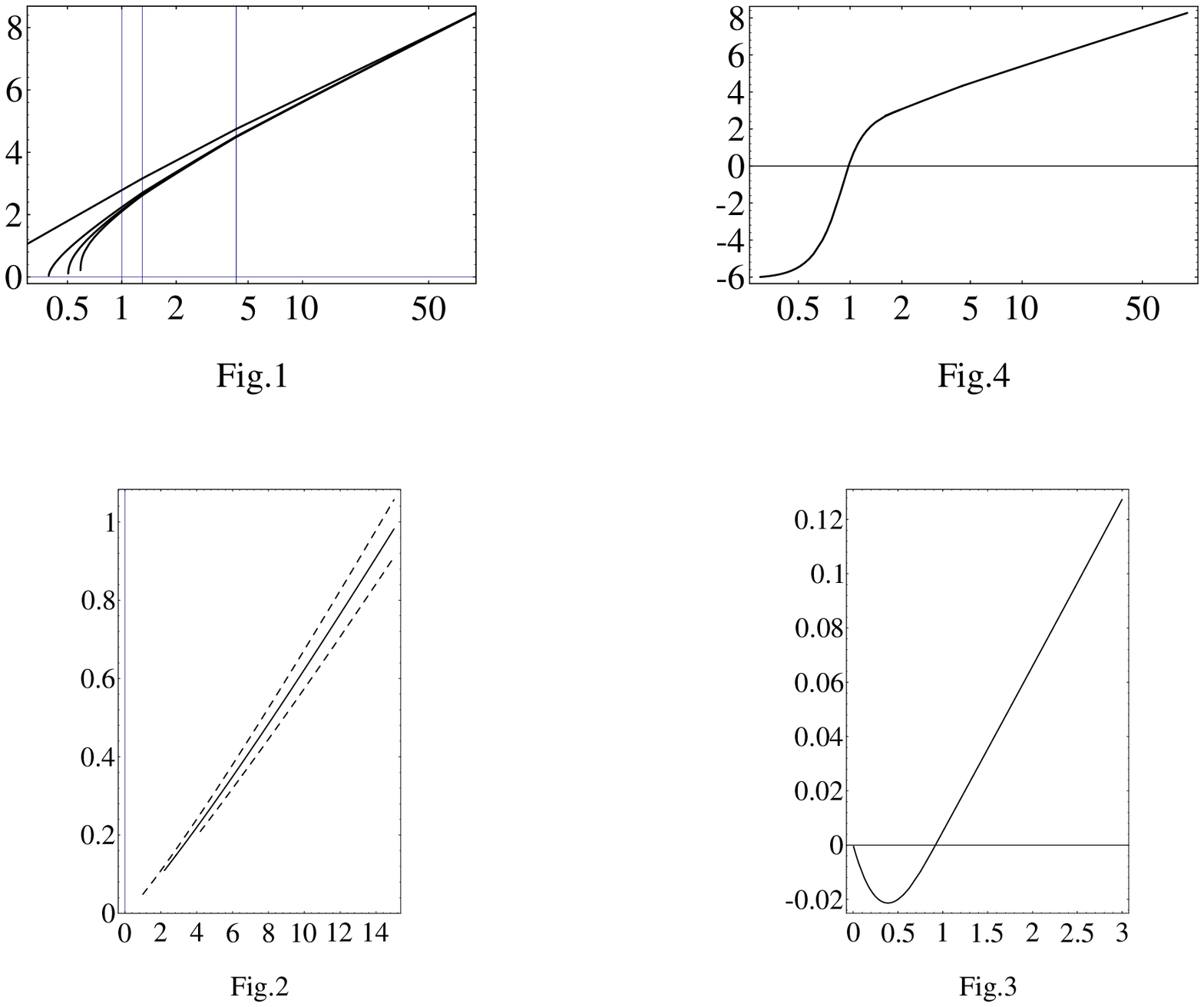}}
%Fig1:
\put(13,82){\llap{$\alpha_s^{-1}$}}
\put(59,56.5){$\mu$,GeV}
\put(26,85){c}
\put(34.3,85){b}
\put(21,67){1}
\put(22.5,61){4}
%Fig4:
\put(77.5,82){\llap{$\alpha_s^{-1}$}}
\put(124.5,56.5){$\mu$,GeV}
%Fig2:
\put(52,-1){$F^2,$GeV$^4$}
\put(20,37){\llap{$U$}}
\put(51,39){$-\sigma$}
\put(51,33.5){$+\sigma$}
%Fig3:
\put(85,37){\llap{$U=-{\cal L}_{eff}$}}
\put(118,-1){$F^2,$GeV$^4$}
\ep
%\caption{}
\ec
\end{figure}
~


\vspace{0cm}
\begin{thebibliography}{99}
\bibitem{1} S.G.Matinyan, G.K.Savvidy, Nucl.Phys. {\bf B 134} (1978) 539;
\bibitem{2} H.Pagels, E.Tomboulis, Nucl.Phys. {\bf B 143} (1978) 485;
\bibitem{3} M.A.Shifman, A.I.Vainstein, V.I.Zakharov, Nucl.Phys. {\bf B 147} (1975) 385;
\bibitem{4} O.V.Tarasov, A.A.Vladimirov, A.Yu.Zharkov, Phys.Lett. {\bf B 93} (1980) 429;
\bibitem{5} T.Ritbergen, J.Vermaseren, S.Larin, hep-ph/9701390;
\bibitem{6} Review of Particle Physics, Part I, Phys.Rev. {\bf D 54} (1996)1.
\bibitem{7} V.V.Vladimirsky, Yad.Fiz. {\bf 59} (1996) 2063.
\end{thebibliography}
\end{document}